\documentclass[twocolumn,showpacs,preprintnumbers,superscriptaddress,amsmath,amssymb]{revtex4}

\usepackage{amsmath}
\usepackage{graphicx}
\usepackage{dcolumn}
\usepackage{bm}

\usepackage{mathrsfs}

\usepackage{color}

\begin{document}

\preprint{1}

\title{Collision system size scan of collective flows in relativistic heavy-ion collisions}

\author{S. Zhang}
\affiliation{Key Laboratory of Nuclear Physics and Ion-beam Application~(MOE), Institute of Modern Physics, Fudan University, Shanghai $200433$, China}

\author{Y. G. Ma\footnote{Author to whom all correspondence should be addressed: mayugang@fudan.edu.cn}}
\affiliation{Key Laboratory of Nuclear Physics and Ion-beam Application~(MOE), Institute of Modern Physics, Fudan University, Shanghai $200433$, China}
\affiliation{Shanghai Institute of Applied Physics, Chinese Academy of Sciences, Shanghai 201800, China}

\author{G. L. Ma}
\affiliation{Key Laboratory of Nuclear Physics and Ion-beam Application~(MOE), Institute of Modern Physics, Fudan University, Shanghai $200433$, China}

\author{J. H. Chen}
\affiliation{Key Laboratory of Nuclear Physics and Ion-beam Application~(MOE), Institute of Modern Physics, Fudan University, Shanghai $200433$, China}

\author{Q. Y. Shou}
\affiliation{Key Laboratory of Nuclear Physics and Ion-beam Application~(MOE), Institute of Modern Physics, Fudan University, Shanghai $200433$, China}

\author{W. B. He}
\affiliation{Key Laboratory of Nuclear Physics and Ion-beam Application~(MOE), Institute of Modern Physics, Fudan University, Shanghai $200433$, China}

\author{C. Zhong}
\affiliation{Key Laboratory of Nuclear Physics and Ion-beam Application~(MOE), Institute of Modern Physics, Fudan University, Shanghai $200433$, China}

\date{\today}

\begin{abstract}
Initial geometrical distribution and fluctuation can affect the collective expansion in relativistic heavy-ion collisions. This effect may be more evident in small system (such as B + B) than in large one (Pb + Pb). This work presents the collision system dependence of collective flows and discusses about effects on collective flows from initial fluctuations in a framework of a multiphase transport model. The results shed light on system scan on experimental efforts to small system physics.
\end{abstract}

\maketitle


\section{Introduction}
Quark-Gluon Plasma (QGP) state was predicted by Quantum chromodynamics (QCD) and could be formed under ultra dense-hot conditions by heavy ion collisions~\cite{QCD-QGP}. This new state of nuclear matter is considered to be produced at the early state of central nucleus-nucleus collisions in experiments~\cite{RHICWhitePaper-STAR,FisrtResultsALICE}, which presents collective motion in partonic level~\cite{PhysRevLett.99.112301}. The properties of QGP are still an open question in heavy-ion collision community, which is not only depending on properties of QCD but also sensitive to initial geometry and dynamical fluctuations. The initial geometrical distribution and fluctuation can remain influence to observables at final state, such as collective flows~\cite{STARv1BES,STARv2BES, STARv3,STARFlowCME,PhysRevLett.121.222301,smallSystemQGPPHENIX}, Hanbury-Brown-Twiss (HBT) correlation~\cite{STAR-HBTBES,PHENIX-HBTBES} and fluctuation~\cite{STAR-FLUCBES}. Some theoretical works~\cite{PartPlane-1,PartPlane-2,PartPlane-3} presented flow/eccentricity analysis methods related to initial geometry fluctuations.  A multi-phase transport (AMPT) model~\cite{AMPTInitFluc-1,AMPTInitFluc-2,AMPTInitFluc-3} demonstrated initial geometry fluctuations of partons created in Au + Au collisions and its effects on elliptic and triangular flow. Initial fluctuation effects have been  also proposed on some observables or physics quantities, such as on collective obeservabels~\cite{NSTSongFlow,NSTWangFlow,GuoCC,XuZW}, conserved quantities~\cite{NSTLuoCQ}, density fluctuations~\cite{NSTKeDensity}, and chiral effects  \cite{ShouQY,WangFQ} etc.

To understand these observables in experiments and QGP phase transition, there are also some open questions for small collision systems (such as C + C or O + O collisions) as well as large collision systems (such as Au + Au or Pb + Pb collisions), (1) how to understand transformation coefficient from initial geometry distribution or fluctuation to momentum distribution at final stage in hydrodynamical mechanism~\cite{NSTSongFlow,HydroKN-1,HydroKN-2,PhysRevC.100.024904}; (2) how to understand similar phenomena for some observables for small systems with high multiplicity  and  large systems~\cite{PhysRevLett.121.222301,smallSystemQGPPHENIX,dAuPHENIXPRL,smallSystemSTAR,smallSystemCMS2013,smallSystemATLAS2013,smallSystemALICE2013,smallSystemATLAS2016,smallSystemALICE2017}; (3) does the matter created in different size of system undergo the similar dynamical process and have similar viscosity ~\cite{vnen_beta_Ntrack201809,vnen_beta_Ntrack201808}?, and the last two questions are closely related. Recently  the small system experiments were proposed for RHIC-STAR~\cite{smallSystemSLHuang} and LHC-ALICE~\cite{smallSystemLHC} to study the initial geometry distribution and fluctuations effects on momentum distribution at final stage. And lots of theoretical works contributed physics explanation and analysis method to this subject~\cite{smallSystem-th-JLN2014,smallSystem-th-SHL2019-1,smallSystem-th-SHL2019-2,smallSystem-th-KW2016,smallSystem-th-XLZ2018,PhysRevC.100.024904,arXiv:1907.03308v1}. By using Trento+v-USPhydro, Ref.~\cite{PhysRevC.100.024904} investigated the response of collective flow to initial geometry asymmetry in small and large systems, and based on Trento+v-USPhydro+DAB-MOD,  Ref.~\cite{arXiv:1907.03308v1} made predictions for system size scan of heavy flavour flow in the LHC energy region. And some detailed introduction and discussion can be found in some recent review works~\cite{smallSsystemReviewEXPLHC,smallSystemReview}.

In this work, a system scan from $^{10}$B + $^{10}$B to $^{208}$Pb + $^{208}$Pb  are studied by using a multi-phase transport (AMPT) model. The collective harmonic flow coefficients ($v_n$, $n$=2,3,4) are calculated, as well as the fourth order linear and nonlinear mode coefficient $v_4^L$ and $v_{4,22}$. The corresponding initial geometry eccentricity coefficients are also presented. It is found that $v_n$ decreases smoothly with the increasing of multiplicity created in the most central collisions of different system as the system size dependence of eccentricity. The response of collective flow to initial geometry asymmetry are discussed and the system size dependence of the response are related to the viscous properties of the system. It suggests to investigate the response for higher order nonlinear mode in a system scan experiment project.
 
\section{A brief introduction to AMPT and algorithm}

In this work, the relativistic heavy-ion collisions are simulated by a multi-phase transport  model~\cite{AMPT2005} with version $2.26t7b$. The initial state of the collisions is described by the Heavy Ion Jet Interaction Generator (HIJING) model~\cite{HIJING-1,HIJING-2} and the melted partons from HIJING interact with each other in the Zhang's Parton Cascade (ZPC) model~\cite{ZPCModel}.  And then the interacting-ceased partons are converted to hadrons by a simple quark coalescence model or the Lund string fragmentation. The hadrons participate in rescattering process through a relativistic transport model~\cite{ARTModel}.  AMPT was successful to describe physics in relativistic heavy-ion collisions for RHIC~\cite{AMPT2005} and LHC~\cite{AMPTGLM2016}, including pion-HBT correlations~\cite{AMPTHBT}, di-hadron azimuthal correlations~\cite{AMPTDiH,WangHai}, collective flow~\cite{STARFlowAMPT,AMPTFlowLHC} and strangeness production~\cite{NSTJinS,SciChinaJinS}.

The hot-dense matter created in collisions expands in longitudinal direction (i.e. always defined by beam direction) as well as transverse direction. In transverse direction, distribution of produced particles in momentum space can be  expanded in azimuthal distribution as~\cite{flowMethod1998},
\begin{eqnarray}
E\frac{d^3N}{d^3p} = \frac{1}{2\pi}\frac{d^2N}{p_Tdp_Tdy}\left(1+\sum_{i=1}^{N}2v_n\cos[n(\phi-\Psi_{RP})]\right),
\label{FlowExpansion}
\end{eqnarray}
where $E$ is the energy, $p_T$ is transverse momentum, $y$ is rapidity, and $\phi$ is  azimuthal angle of the particle. $\Psi_{RP}$ is reaction plane angle. The Fourier coefficients, $v_n (n = 1, 2, 3, ...)$, characterize the collective flows of different orders in azimuthal anisotropies.

The collective flow is driven from the initial anisotropy in geometry space. To investigate transformation from geometry to final momentum space, the initial geometry eccentricity coefficients $\varepsilon_{n}$ can be calculated from the participants via~\cite{AMPTInitFluc-1,PartPlane-1,PartPlane-2,PartPlane-3,HydroKN-1,vnen_beta_Ntrack201808},
\begin{eqnarray}
\mathrm{\mathcal{E}_n  \equiv \varepsilon_n e^{in\Phi_n} \equiv 
  - \frac{\langle r^{n}e^{in\phi_{Part}}\rangle}
           {\langle r^{n}\rangle}},    
\label{EpsilonPPDef}
\end{eqnarray}
where, $r$=$\sqrt{x^2+y^2}$ and $\phi_{Part}$ are coordinate position and azimuthal angle of initial participants in the collision zone in the recentered coordinates system ($\langle x\rangle$=$\langle y\rangle$=0). $\Phi_n$ is the initial participant plane and $\varepsilon_n$=$\langle|\mathcal{E}_n|^2\rangle^{1/2}$. The bracket $\langle\rangle$ means the average over the transverse position of all participants event by event. Note that for the definition of  eccentricity coefficients $\varepsilon_{n}$, $r^{2}$ weight was alternative and it was discussed in Refs.~\cite{HydroKN-1,PartPlane-3}.

Two particle correlation (2PC) method with $\Delta\eta$ gap is usually employed to calculate the collective flow coefficients in theoretical analysis and experimental measurements~\cite{TwoPartCorrRap-1,TwoPartCorrRap-2,TwoPartCorrRap-3,TwoPartCorrRap-4,TwoPartCorrRap-5} . In this work we adopted the 2PC-method introduced in Ref.~\cite{TwoPartCorrRap-1} to calculate transverse momentum $p_T$ and centrality dependence of the collective flow.

The Q-cumulant method~\cite{EP-QC-Method-1,QC-method-1,AMPTInitFluc-1,AMPTInitFluc-2,HydroKN-1,vnen_beta_Ntrack201808} is also popular in flow coefficients analysis. The complex flow vectors~\cite{PLB2015_nonlinear2,vnen_beta_Ntrack201808} is defined by $V_n \equiv v_ne^{in\Psi_n} \equiv \{e^{in\phi}\}$, $v_n=\langle|V_n|^2\rangle^{1/2}$, where $\phi$ is azimuthal angle of final particles, $v_n$ and $\Psi_n$ is the $n$th order flow coefficients and azimuthal direction of the event, $\{...\}$ denotes the average over all final particles in each event. 

For the higher-order collective flow coefficients ($n>3$), the nonlinear mode couplings derived from lower-order collective flow coefficients should be taken into account except the linear response related to eccentricity, which was discussed in Refs.~\cite{PhysRevC2012_nonlinear1,PLB2015_nonlinear2,vnen_beta_Ntrack201808}. Here we employ the formulas of the fourth order linear-mode, nonlinear-mode flow and geometry coefficients suggested in Ref.~\cite{vnen_beta_Ntrack201808}, i.e. $v_{4,22}$  $\approx$ $\langle v_4\cos(4\Psi_4-4\Psi_2)\rangle$, $v_4^L$=$\sqrt{v_4^2-v_{4,22}^2}$,  $\mathcal{E}_4^{L}$=$\mathcal{E}_4+\frac{3\langle r^2\rangle^2}{\langle r^4\rangle}\mathcal{E}_2^2$, and $\varepsilon_{4,22}$ = $\sqrt{\langle\epsilon_2^4}\rangle$. 

From hydrodynamics viewpoint, the relationship between initial geometry eccentricity coefficients and flow coefficients can be described by $v_n \propto \varepsilon_n$,  (n = 2, and 3)~\cite{NSTSongFlow,HydroKN-1,HydroKN-2}.
The response of $v_n$ to $\varepsilon_n$ showed the efficiency of the transformation from initial geometry properties to final momentum space in heavy-ion collisions. For higher-order initial geometry eccentricity coefficients and flow coefficients, the relationship can be described by linear and nonlinear-mode~\cite{PhysRevC2012_nonlinear1,PLB2015_nonlinear2,vnen_beta_Ntrack201808}, $v_n^L$ $\propto$ $\varepsilon_n^L$ and $v_{n,ij} \propto \varepsilon_{n,ij}$. Hydrodynamics with viscous corrections gives the acoustic scaling of anisotropic flow in shape-engineered events~\cite{vnen_beta_Ntrack201809,vnen_beta_Ntrack201808,vnen_beta_Ntrack2013,vnen_beta_Ntrack2011}, 
\begin{eqnarray}
v_n^{L}/\varepsilon_n^{L} \propto \exp\left(-n^2\beta\left<N_{track}\right>^{-1/3}\right),\\
v_{n,ij}/\varepsilon_{n,ij} \propto \exp\left(-(i^2+j^2)\beta\left<N_{track}\right>^{-1/3}\right),
\label{vnenNpartScaling}
\end{eqnarray}
where $L$ for $n>3$, the parameter $\beta$ is related to ratio of shear viscosity ($\eta$) over entropy density ($s$), namely $\beta\propto\eta/s$ and $\left<N_{track}\right>$ average number of particles created in the collisions with kinetic windows (always in mid-rapidity ($|y|<1$) and $0.2<p_T<4$ GeV/c).

\section{Results and discussion}

By using AMPT model, a system scan simulation is performed in this work involving the most central collisions (i.e. impact parameter $b$ is set to zero) of $^{10}$B + $^{10}$B, $^{12}$C + $^{12}$C, $^{16}$O + $^{16}$O, $^{20}$Ne + $^{20}$Ne, $^{40}$Ca + $^{40}$Ca, $^{96}$Zr + $^{96}$Zr and $^{208}$Pb + $^{208}$Pb systems,  at center of mass energy $\sqrt{s_{NN}}=$6.73 TeV. The generated event numbers are  presented in Table~\ref{tab:evtNum}. Via the introduced flow analysis methods, the harmonic flow coefficients are calculated in these collision systems under the kinetic windows, transverse momentum $0.2<p_T<3$ GeV/$c$ and rapidity $|y|<1.0$.

\begin{table*}
\caption{ \label{tab:evtNum} Collision system and number of event for each system. }
\begin{ruledtabular}
\begin{tabular}{llllllll}
system & $^{10}$B+$^{10}$B & $^{12}$C+$^{12}$C & $^{16}$O+$^{16}$O & $^{20}$Ne+$^{20}$Ne & $^{40}$Ca+$^{40}$Ca & $^{96}$Zr+$^{96}$Zr & $^{208}$Pb + $^{208}$Pb\\
 \hline
 event number & 2.3$\times10^{5}$ & 4.2$\times10^5$ & 2.4$\times10^5$ & 1.9$\times10^5$ & 4.9$\times10^4$ & 1.4$\times10^4$ & 3.5$\times10^3$
\end{tabular}
\end{ruledtabular}
\end{table*}

\begin{figure}[htb]
\includegraphics[angle=0,scale=0.44]{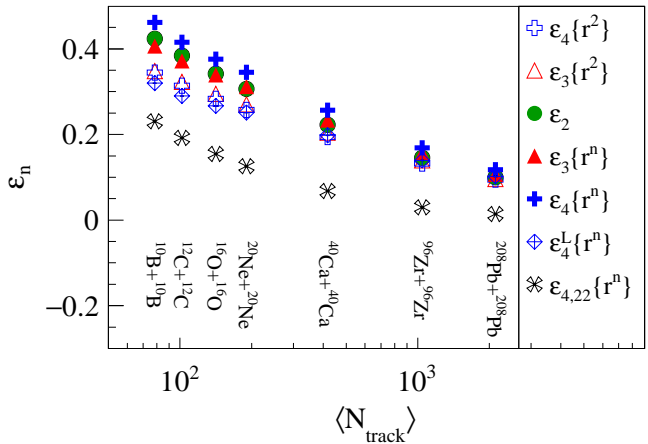}
\caption{Eccentricity coefficients for the most central collision events in $^{10}$B + $^{10}$B, $^{12}$C + $^{12}$C, $^{16}$O + $^{16}$O,  $^{20}$Ne + $^{20}$Ne, $^{40}$Ca + $^{40}$Ca, $^{96}$Zr + $^{96}$Zr and $^{208}$Pb + $^{208}$Pb  at center of mass energy $\sqrt{s_{NN}} = $6.73 TeV.
\label{fig:en}
}
\end{figure}

The initial geometry eccentricity coefficients $\varepsilon_n$ ($n$ = 2, 3, 4) are calculated by using Eq.~(\ref{EpsilonPPDef}) and the nonlinear-mode eccentricity coefficients are also calculated, as shown in figure~\ref{fig:en}. The eccentricity coefficients $\varepsilon_n$ ($n$=2,3,4) are all smoothly decreasing with the increasing of size of collision systems ($\left<N_{track}\right>$) from $^{10}$B + $^{10}$B collisions to $^{208}$Pb + $^{208}$Pb collisions. The fourth order linear-mode $\varepsilon_{4,22}$ and nonlinear-mode coefficients $\varepsilon_4^L$ also presented the similar system size dependence. The initial geometry eccentricity coefficients $\varepsilon_n$ ($n$ = 3, 4) are calculated with $r^2$ and $r^n$ weight separately. It can be seen from the results in smaller system that $\varepsilon_n$ (n=3,4) is higher with $r^{n}$ weight than with $r^2$ weight, which is mainly due to the weight from periphery of the initial system. However, the different $r$ wights give the comparable value of $\varepsilon_n$ (n=3,4) for larger systems. For smaller systems by $r^2$ weight it shows $\varepsilon_2>\varepsilon_3\sim\varepsilon_4$ and by $r^n$ weight $\varepsilon_4>\varepsilon_2\sim\varepsilon_3$ and $\varepsilon_2>\varepsilon_3>\varepsilon_4^L$. In the following calculations,  we only use eccentricity coefficients calculated by $r^n$ weight. This system size dependence of eccentricity coefficients indicates that geometrical fluctuation is more significant in small system than in large system in the most central collisions ($b$ set zero). 

\begin{figure*}[htb]
\includegraphics[angle=0,scale=0.9]{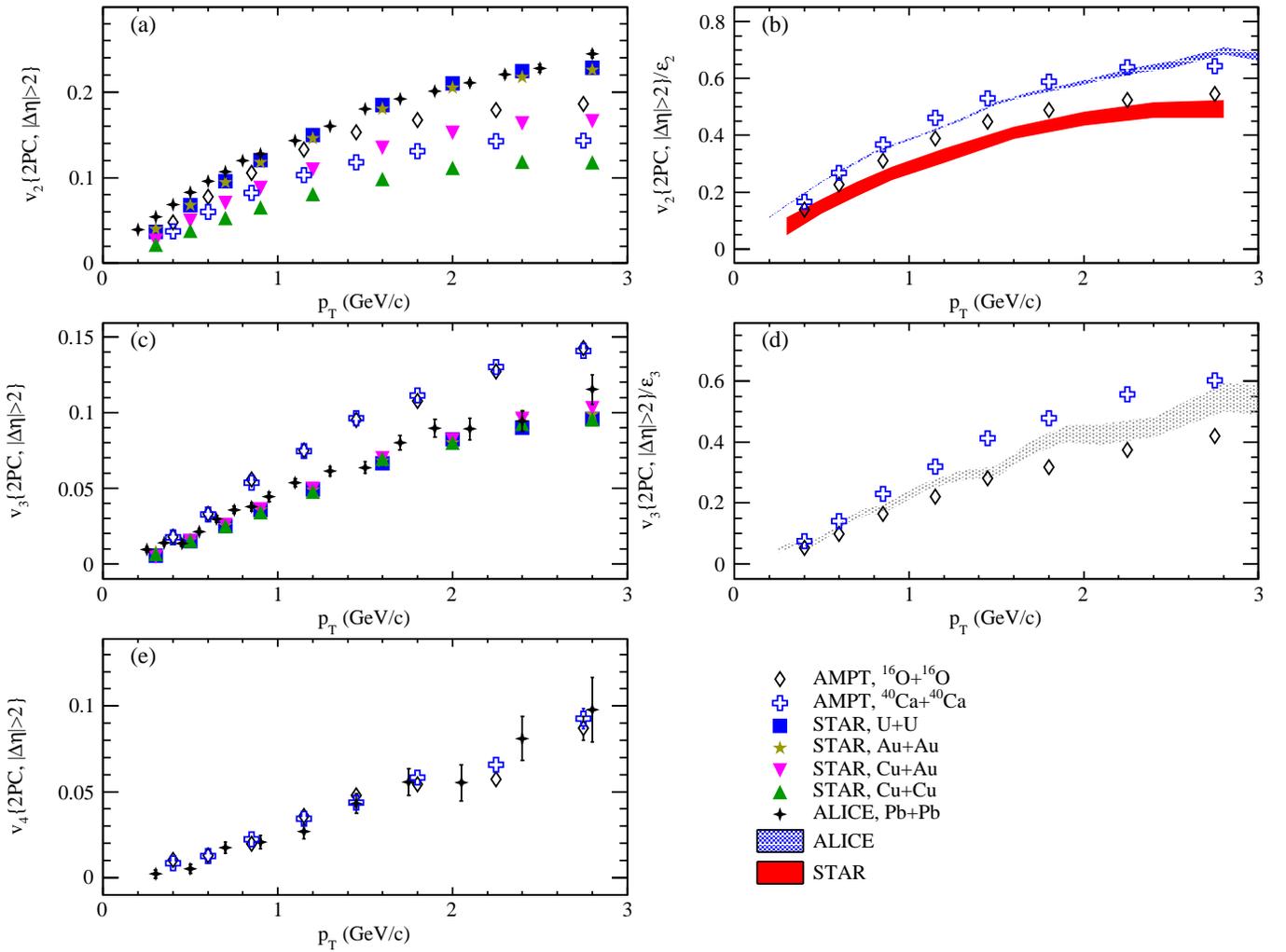}
\caption{Left columns: Collective flows $v_2$,  $v_3$ and $v_4$ as a function of $p_T$ for the most central collision events in $^{10}$B + $^{10}$B, $^{12}$C + $^{12}$C, $^{16}$O + $^{16}$O,  $^{20}$Ne + $^{20}$Ne, $^{40}$Ca + $^{40}$Ca, $^{96}$Zr + $^{96}$Zr and $^{208}$Pb + $^{208}$Pb at center of mass energy $\sqrt{s_{NN}} = $6.73 TeV via two-particle correlation method. Right columns: the ratios of $v_n/\epsilon_n$ (n=2,3) as a function of $p_T$.
\label{fig:vn-pT-withExp}(color online)
 }
\end{figure*}

The collective flows $v_{n}$ (n=2,3,4) as a function of transverse momentum $p_T$ are shown in panel (a), (c) and (e) in figure~\ref{fig:vn-pT-withExp} for $^{16}$O+$^{16}$O ($\langle N_{track}\rangle$=141) collisions and $^{40}$Ca+$^{40}$Ca ($\langle N_{track}\rangle$=418) collisions at$\sqrt{s_{NN}}$ = 6.73 TeV by using two-particle correlation method with $|\Delta\eta|>1$. Also,  the $p_T$ dependence of $v_n/\epsilon_n$ (n=2,3) is also calculated and presented in panel (b) and (d) respectively. The results are also compared with experimental measurements by the RHIC-STAR data~\cite{STARSmallSys} in A+A (U+U, Au+Au, Cu+Au, Cu+Cu) collisions ($\left<N_{track}\right>$=140, $|\eta|<1.0$) at top RHIC energy by two particle correlation method with $|\Delta\eta|>0.7$ and the LHC-ALICE data~\cite{ALICESmallSys}  in Pb+Pb collisions (centrality 30-40\%, $\left<N_{track}\right>$=426, $|\eta|<0.8$) at $\sqrt{s_{NN}}$ = 2.76 TeV by two particle correlation method with $|\Delta\eta|>1$. Note that $\langle N_{track}\rangle$ in $^{16}$O+$^{16}$O ($^{40}$Ca+$^{40}$Ca) collisions is approximate to that in A+A collisions from RHIC-STAR (ALICE) experiment.
It is found that the collective flows increase with $p_T$ which are similar to those from experiments. In $^{16}$O+$^{16}$O collisions the elliptic flows are close to those  in Cu+Au collisions from STAR experiments, and the triangular flows are higher than that from experiments. While in $^{40}$Ca+$^{40}$Ca collisions the elliptic flows are lower than that from ALICE experiment, triangular flows are higher than ALICE experiment results and quadrangular flows are similar to experiment results. 
These collision systems have different eccentricity and this comparison can not give more information about transferring asymmetry from geometry space to momentum space. Panel (b) and (d) in figure~\ref{fig:vn-pT-withExp} give the $v_{n}/\epsilon_n$ (n=2,3), respectively. $v_{n}/\epsilon_n$ (n=2,3) are larger in Ca+Ca collisions than in O+O collisions by AMPT model. Since the $v_2/\epsilon_2$ from the STAR collaboration in U+U, Au+Au, Cu+Au, Cu+Cu collisions drops in one group, the plot gives a band with the maximum uncertainties from Ref.~\cite{STARSmallSys}. $v_2/\epsilon_2$ and $v_3/\epsilon_3$ from the ALICE collaboration in Pb+Pb collisions with centrality 30-40\% from Ref.~\cite{ALICESmallSys}. It is found the $p_T$ dependence of $v_{n}/\epsilon_n$ (n=2,3) in Ca+Ca (O+O) collisions by AMPT model are consistent with those from the ALICE (STAR) and display an obvious system size dependence of $v_n/\epsilon_n$.

\begin{figure*}[htb]
\includegraphics[angle=0,scale=0.90]{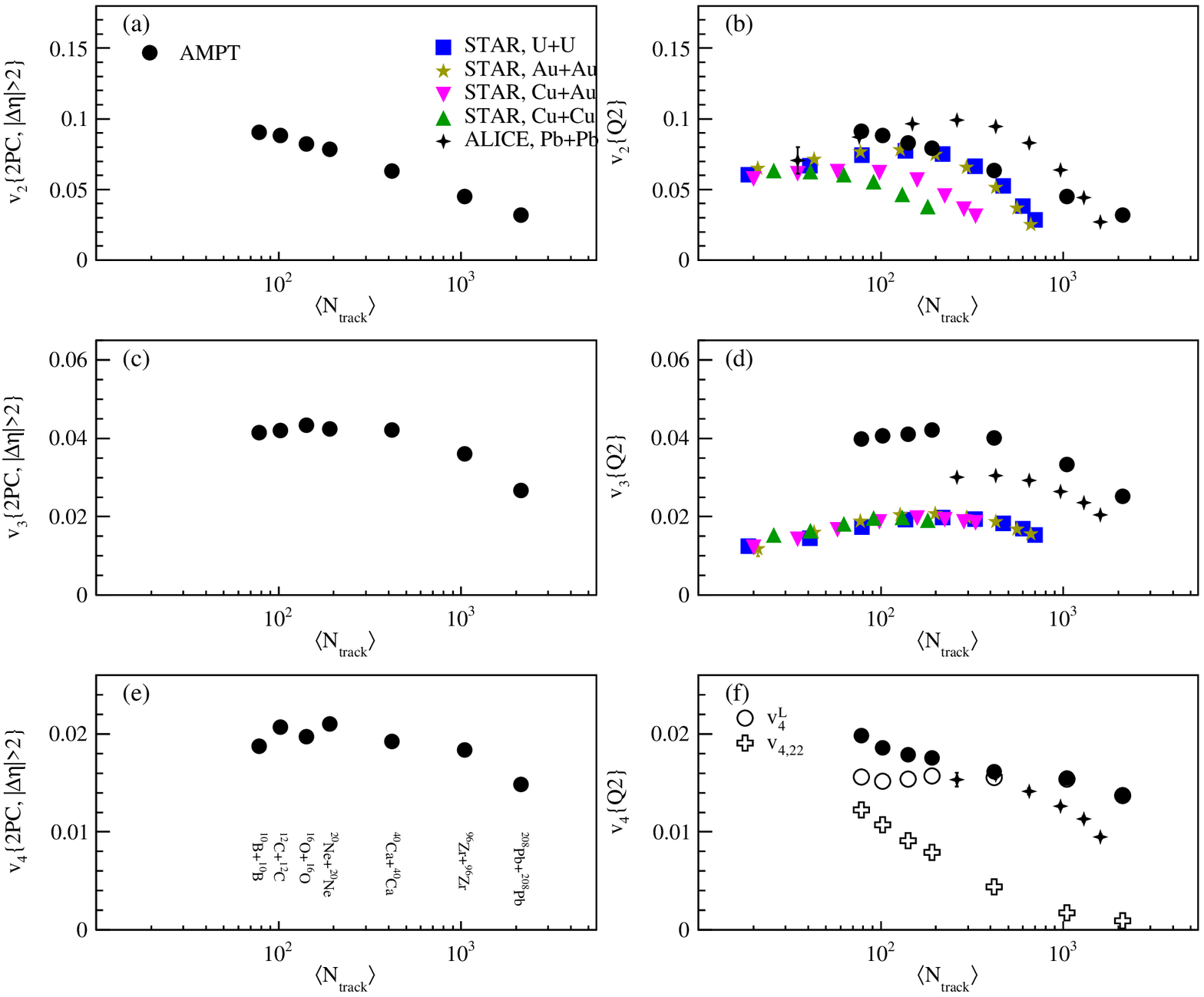}
\caption{Collective flows calculated by two-particle correlation method and cumulant method in different collision systems.
\label{fig:vn-Ntrack}(color online)
 }
\end{figure*}

\begin{figure*}[htb]
\includegraphics[angle=0,scale=0.90]{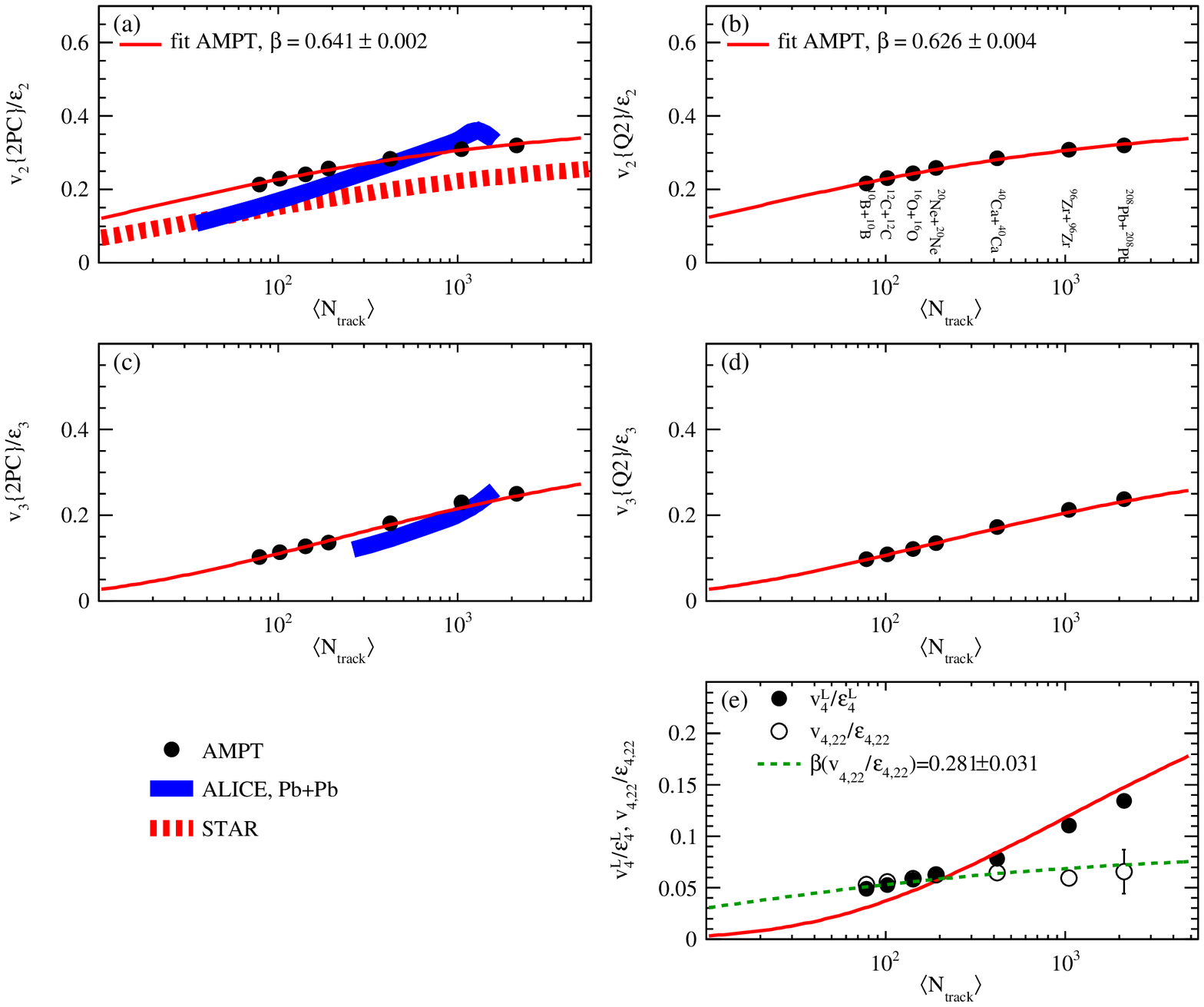}
\caption{Ratio of different order collective flows to initial geometry eccentricities for different collision systems.
\label{fig:vnen-scalingNtrack}(color online)
 }
\end{figure*}

The $p_T$ integrated collective flows $v_2$,  $v_3$ and $v_4$ are calculated in the above introduced collision systems at center of mass energy $\sqrt{s_{NN}}$ = 6.73 TeV and sown in figure~\ref{fig:vn-Ntrack}, via different flow analysis methods, i.e. panels (a), (c) and (e) represent two-particle correlation with $\Delta\eta>1$, panels (b), (d) and (f) represent 2-particle cumulants.  The fourth order nonlinear-mode $v_{4,22}$ and linear-mode $v_{4}^{L}$ calculated by 2-particle cumulants method are also shown in panel (f) in figure~\ref{fig:vn-Ntrack}.
The collective flows  $v_2$,  $v_3$ and $v_4$ decrease with the increasing of collision system size from $^{10}$B + $^{10}$B to $^{208}$Pb + $^{208}$Pb at the most central collisions, as the system size dependence of eccentricity displays, by using AMPT model.   $v_{4,22}$ presents the similar system size dependence as $v_4$ and this results in that $v_4^L$ shows a more flat trend with the increasing of system size. The experimental results are from the STAR data~\cite{STARSmallSys} in U+U, Au+Au, Cu+Au, C+Cu collisions and from ALICE data~\cite{ALICESmallSys} in Pb+Pb collisions. Two particle correlation method with $\Delta\eta$ gap was adopted in the experimental flow analysis. The STAR analysis use cuts of $0.2<p_T<4$ GeV/$c$, $|\eta|<1.0$ and $|\Delta\eta|>0.7$, ALICE cuts of $0.2<p_T<5$ GeV/$c$ and $|\eta|<0.8$ and $|\Delta\eta|>1.0$. The elliptic flow $v_2$ and triangular $v_3$ in Au+Au and U+U collisions increase and then decrease with the increasing of $\left<N_{track}\right>$. $v_2$ in Pb+Pb collisions also presents the similar $\left<N_{track}\right>$ dependence. $v_2$ from AMPT presents a more linear $\left<N_{track}\right>$ dependence and is different from those of experiments, $v_3$ and $v_4$ give the similar $\left<N_{track}\right>$ dependence trend as experimental results demonstarte.


To further investigate the system size dependence of momentum asymmetry from initial geometry asymmetry, the response of collective flow to initial geometry asymmetry, $v_2/\varepsilon_2$, $v_3/\varepsilon_3$, $v_4^L/\varepsilon_4^L$, $v_{4,22}/\varepsilon_{4,22}$, are calculated and shown in figure~\ref{fig:vnen-scalingNtrack}. Panel (a) and  (c) are the results with flow coefficients via two-particle correlation method and  panels (b), (d) and (e) with that via two-particle cumulants. $v_2/\varepsilon_2$, $v_3/\varepsilon_3$, $v_4^L/\varepsilon_4^L$, and  $v_{4,22}/\varepsilon_{4,22}$  increase with collision system size and different flow analysis methods present the same system dependence with similar values.  The upwards trend of system size dependence of the response indicates that the transferring efficiency from initial geometry asymmetry to final momentum is higher in large size collision systems than in small ones, on other words the higher multiplicity the system has, the higher transferring efficiency the evolution of system would get. These results are similar to those from experiments~\cite{STARSmallSys,ALICESmallSys} and more closer to ALICE results in Pb+Pb collisions as a function of $\sqrt{s_{NN}}$.

The extracted $\beta$ by equation~(\ref{vnenNpartScaling}) from $\left<N_{track}\right>$ dependence of the response of $v_n$ to $\varepsilon_n$ could provide information if the collision systems present similar shear viscous properties, in other words if the collision systems exhibit the similar QGP fluid friction  and undergoes similar dynamical process. The lines on figure~\ref{fig:vnen-scalingNtrack} show simultaneously fitting for $v_n/\varepsilon_n$ ($n$=2,3) and $v_4^L/\varepsilon_4^L$ by equation~(\ref{vnenNpartScaling}). It give $\beta$ = $0.641\pm0.002$ for two-particle correlation method and $\beta$ = $0.626\pm0.004$ for two-particle cumulant method. The fitting results are similar to those ($\beta$=$0.82\pm0.02$) extracted by fitting $v_2/\epsilon_2$ in small and large collision system at top RHIC energies by RHIC-STAR~\cite{STARSmallSys}.  It is interesting that the fitting to $v_{4,22}/\varepsilon_{4,22}$ give lower value of $\beta$ = $0.281\pm0.031$. In Ref.~\cite{PhysRevC.83.034904} the ratio of shear viscosity to entropy was estimated to be 0.273 in Pb+Pb collisions at $\sqrt{s_{NN}}$=2.76 TeV. $\beta$ extracted by using equation~(\ref{vnenNpartScaling}) from the response of $v_n$ to $\varepsilon_n$ is higher than that in Ref.~\cite{PhysRevC.83.034904}, and $\beta$ from $v_{4,22}/\varepsilon_{4,22}$ is closer to that. Note that this maybe result from different methods and difference parameter setting in previous AMPT result ~\cite{PhysRevC.83.034904} ($\sqrt{s_{NN}}$ = 2.76 TeV, $\mu$=3.2 $fm^{-1}$) and in this work ($\sqrt{s_{NN}}$ = 6.73 TeV, $\mu$=2.3$fm^{-1}$), here $\mu$ is the screening mass in the partonic matter. These results indicate that it should be investigated in experiment for energy dependence of  $\beta$ and the response of collective flow to initial asymmetry for higher-order nonlinear mode. 


\section{Summary}
The collective flow harmonic coefficients are calculated and presented at $\sqrt{s_{NN}}$ = 6.73 TeV for the most central collision systems from small one to large one, namely $^{10}$B + $^{10}$B, $^{12}$C + $^{12}$C, $^{16}$O + $^{16}$O, $^{20}$Ne + $^{20}$Ne, $^{40}$Ca + $^{40}$Ca, $^{96}$Zr + $^{96}$Zr and $^{208}$Pb + $^{208}$Pb collisions. From these results, it is found that collective flows show smooth changing trend with the increasing of the collision system size and is sensitive to initial geometry eccentricities. The response of collective flows to initial geometry asymmetries, namely $v_2/\varepsilon_2$, $v_3/\varepsilon_3$, $v_4^L/\varepsilon_4^L$, $v_{4,22}/\varepsilon_{4,22}$, are also calculated and seems sensitive to system size (or multiplicities). With aid of hydrodynamics with viscous corrections, the acoustic scaling of anisotropic flow in shape-engineered events is performed to the system size dependence of collective flow. The parameter $\beta$ related to shear viscosity over entropy density ratio seems consistent with that from experiments, but $\beta$ from $v_{4,22}/\varepsilon_{4,22}$ is lower than those from $v_2/\varepsilon_2$, $v_3/\varepsilon_3$ and $v_4^L/\varepsilon_4^L$. The system scan experiment is therefore  proposed to systematically explore the effects from initial geometry fluctuations, and then the transformation efficiency from initial geometry to final momentum could be studied.

We are grateful for discussions with Profs. J. Y. Jia (Stony Brook University), C. M. Ko (Texas A\&M University), Z. W. Lin (East Carolina University), Aihong Tang (BNL), and Constantin Loizides (Oak Ridge National Laboratory). This work was supported in part by the National Natural Science Foundation of China under contract Nos. 11875066,  11890714, 11421505 and 11775288, National Key R\&D Program of China under Grant No. 2016YFE0100900 and 2018YFE0104600, the Key Research Program of Frontier Sciences of the CAS under Grant No. QYZDJ-SSW-SLH002, and the Key Research Program of the CAS under Grant NO. XDPB09.


\bibliography{mybibfile}

\end{document}